\pdfoutput=1
\documentclass[apjl]{latex/emulateapj}

\usepackage{graphicx}
\usepackage{xspace}
\usepackage{amsmath}

\shorttitle{$\tau$ of clusters}
\shortauthors{Battaglia}

\newcommand{\be}{\begin{equation}}
\newcommand{\ee}{\end{equation}}
\newcommand{\bea}{\begin{eqnarray}}
\newcommand{\eea}{\end{eqnarray}}
\newcommand{\rmn}{\mathrm}
\newcommand{\te}{\tau}
\newcommand{\st}{\sigma_\mathrm{T}}
\newcommand{\me}{m_\mathrm{e}}
\newcommand{\nume}{n_\mathrm{e}}
\newcommand{\dd}{\mathrm{d}}
\newcommand{\elct}{\mathrm{e}}
\newcommand{\tauy}{\langle \tau \rangle-\langle y \rangle}
\newcommand{\tauM}{\langle \tau \rangle-M}

\newcommand{\systa}{\sigma_\mathrm{sys}}
\begin{document}

\title{The Tau of Galaxy Clusters}
\author{N.~Battaglia$^{1,\star}$}

\altaffiltext{1}{Dept. of Astrophysical Sciences, Princeton University, Princeton, NJ 08544, USA}
\altaffiltext{$\star$}{nbatta@astro.princeton.edu}

\begin{abstract}

The recent emergence of detections of the kinetic Sunyaev-Zel'dovich (kSZ) effect through cross-correlation techniques is encouraging for the prospects of future cosmic microwave background (CMB) experiments.
Extracting information on the large-scale velocity fields and constraining cosmological parameters from such kSZ measurements requires an understanding of the optical depth to CMB photons through halos.
Using cosmological hydrodynamic simulations we find that there exists a low-scatter relation between the optical depth and thermal Sunyaev-Zel'dovich (tSZ) signal of halos within a physical aperture.
We propose that such a relation can be used to break the degeneracy between optical depth and line-of-sight velocity in kSZ measurements. 
The limiting factors in our proposal are systematic uncertainties associated with the sub-grid physics models in the simulations, which we calculate to be less than 10 percent.
We discuss future observational measurements that could potentially be used to mitigate the systematic uncertainties in this scaling relation.

\end{abstract}


\section{Introduction}

The intracluster medium (ICM) is optically thin to the CMB radiation. However, a small fraction of CMB photons are scattered or doppler shifted as they travel through the Universe by free electrons in the ICM or the intergalactic medium (IGM). These interactions attenuate or cause secondary anisotropies in the initial CMB radiation, which are known as the thermal and kinetic Sunyaev-Zeldovich effects \citep{SZ1970,SZ1972}. The tSZ effect is the result of CMB photons Compton up-scattering off free electrons. This effect imprints a unique spectral distortion in the CMB blackbody that is a decrement in thermodynamic temperature at frequencies below 217 GHz and excess at higher frequencies. The kSZ is the result of CMB photons Compton-scattering off free electrons that have a non-zero peculiar velocity with respect to the CMB rest frame. As a result we observe a small shift in CMB temperature in the direction of those free electrons. The magnitudes of the tSZ and the kSZ are proportional to the integrated electron pressure and momentum along the line-of-sight, respectively. Thus, they both depend on the optical depth, which is difficult to infer directly on an object-by-object basis using the CMB\citep[e.g.,][]{Hall2014} or through combining CMB and X-ray observations \citep[e.g.,][]{Sehgal2005}.

If the optical depths of galaxies and galaxy clusters are sufficiently understood then the kSZ measurements are potentially powerful observational probes of the peculiar velocities of these systems.
Measurements of these peculiar velocities can provide further constraints on modified gravity models, the dark energy equation of state, and the sum of neutrino masses \citep[e.g.,][]{Dedeo2005,HM2006,BK2007,BK2008,KS2013,Mueller2015a,Mueller2015b,ALBF2016}. Additionally, the kSZ signal is an unbiased tracer of all the ionized baryons in the Universe \citep{HM2008,HM2009,Ho2009,F16}. Thus, it can be used to conduct a baryon census at low redshift and probe whether in fact there are {\it missing} baryons \citep{FHP1998,Bregman2007}.

The first detection of the kSZ was by \citet{Hand2012} using data from the Atacama Cosmology Telescope \citep[ACT;][]{ACT} and galaxy catalogs from the Sloan Digital Sky Survey III, Baryon Oscillation Spectroscopic Survey \citep{BOSS2011}.
Since this initial detection several other detections have followed from multiple CMB experiments, cross-correlating with other galaxy catalogs using various estimators \citep{PlanckkSZ,Schaan2016,Hill2016,SPTkSZ2016,FDB2016},
as well as from one galaxy cluster \citep{Sayers2013}. Forecasts for kSZ measurements from high-resolution CMB experiments such as AdvACT \citep{AdvACT}, the South Pole Telescope-3G \citep[SPT-3G;][]{SPT3G}, and the proposed CMB Stage 4 (CMB-S4) experiment show substantial improvements \citep{Flender2016,F16} over the current measurements. If kSZ measurements are to fulfill their full cosmological constraining power potential then an understanding of the systematics associated with the optical depth of galaxies and clusters is paramount. 

In this work we propose a method for inferring the optical depth from measurements of the tSZ signal.
We use cosmological hydrodynamic simulations to quantify the properties of the relations we explored.
In Sections \ref{sec:thry} and \ref{sec:sims} we provide a theoretical basis and describe the simulations used in this work.
Section \ref{sec:rels} presents the observable scaling relations for the optical depth and estimates the systematic uncertainties associated with them.
We discuss in Section \ref{sec:disc} possible observational techniques that could reduce the systematic uncertainties associated with the model of sub-grid physics in the simulations and conclude our findings in Sec. \ref{sec:con}.
Additionally, in the Appendix we provide a fitting function for the density profile of halos.

\section{Projected cluster properties}
\label{sec:thry}

The observable amplitudes of the kSZ and tSZ effects are proportional to the line-of-sight momentum and pressure, respectively. The fractional shift in the CMB temperature from the kSZ is sensitive to the combination of optical depth and peculiar velocity, ${\bf \hat{v}_\rmn{p}}$ of each halo along the line-of-sight, 

\be
\frac{\Delta T}{T_\rmn{CMB}} = \frac{\st}{c} \int_\rmn{LOS} e^{-\tau} \nume {\bf \hat{v}_\rmn{p}} \dd l, 
\ee

\noindent where $\st$ is the Thompson cross-section, $c$ is the speed of light, $\nume$ is the electron number density, $T_\rmn{CMB}$ is the CMB temperature, and $\dd l$ is the integral along the line-of-sight (LOS). The optical depth, $\tau$, through any given line-of-sight is defined as

\be
\te = \st \int_\rmn{LOS} \nume \dd l.
\label{eq:tau}
\ee

\noindent We consider the optical depth as a {\it predictable} quantity that will allow us to find the peculiar velocity field. The velocity field is the basis of the powerful role kSZ observations can play in cosmology \citep[e.g.,][]{Mueller2015a,Mueller2015b,ALBF2016}. Similarly, the spectral distortion caused by the tSZ in the observed CMB temperature is a function of frequency $\nu$ and the Compton-$y$ parameter: 

\be
\frac{\Delta T(\nu)}{T_\rmn{CMB}} = f(\nu) y, 
\ee

\noindent where $f(\nu) = x\,\rmn{coth}(x/2) - 4$, $x = h\nu / (k T_\rmn{CMB})$, $h$ is the Planck constant, and $k$ is the Boltzmann constant. Note that we neglected relativistic corrections to the tSZ spectral function $f(\nu)$ \citep[e.g.,][]{Nozawaetal2006}. The Compton-y parameter can be defined similarly to Equation \ref{eq:tau},

\bea
y &=& \frac{\st}{\me c^2} \int_\rmn{LOS} \nume kT_\rmn{e} \dd l, \nonumber \\
&=& \frac{\st}{\me c^2} \int_\rmn{LOS} P_\rmn{e} \dd l,
\label{eq:y}
\eea

\noindent where $\me$ is the electron mass, $T_\rmn{e}$ is the electron temperature, and $P_\rmn{e}$ is the electron pressure. Here we have ignored $T_\rmn{CMB}$, since $T_\rmn{e} \gg T_\rmn{CMB}$.

We consider that on average halos have spherically symmetric profiles $\nume(r)$ and $P_\rmn{e}(r)$. Fitting formulae for $\nume(r)$ are provided in Appendix \ref{sec:fit} and the equivalent formulae for $P_\rmn{e}(r)$ are provided in \citet{BBPS2}. The observed signals, integrated within an aperture, corresponds to cylindrical integrals. We project these observable cylindrical quantities for Equations \ref{eq:tau} and \ref{eq:y} such that 

\be
\mathcal{T}_{\rmn{cyl}} = \st \int \dd \theta^2 \int_\rmn{LOS} \nume \left(\sqrt{l^2 + d^2_A(z)|\theta|^2 } \right) \dd l,
\label{eq:tauproj}
\ee

\noindent and 

\be
Y_\rmn{cyl} = \frac{\st}{\me c^2}  \int \dd \theta^2  \int_\rmn{LOS} P_\rmn{e} \left(\sqrt{l^2 + d^2_A(z)|\theta|^2 } \right) \dd l,
\label{eq:yproj}
\ee

\noindent here $r^2 = l^2+d^2_A(z)|\theta|^2$, $\theta = |\theta|$ is the angular distance from the cluster center in the plane of the sky, and $d_A(z)$ is the angular diameter distance to redshift $z$. 

In the limit where the angular aperture is large, i.e. several viral radii, then the Equations \ref{eq:tauproj} and \ref{eq:yproj} simplify to spherically integrated quantities modulo a factor of $1/d^2_A(z)$, which we define as $\mathcal{T}$ and $Y$. Working in this limit we rewrite Equation \ref{eq:tauproj} as

\bea
\mathcal{T} &=& \st \int^{R_{\Delta}}_0 \nume(r) \dd V \nonumber \\ 
&=& \st N_{\rmn{e},\Delta} \nonumber \\
&=& \st x_\rmn{e} X_\rmn{H} (1 - f_\star)f_\rmn{b}f_\rmn{c} \frac{M_\Delta}{\mu m_p},
\label{eq:tauobs}
\eea

\noindent where $x_\rmn{e}$ is the electron fraction defined as $x_\rmn{e} = (X_\rmn{H} + 1) / (2X_\rmn{H})$,
$X_\rmn{H}$ is the primordial hydrogen mass fraction ($X_\rmn{H} = 0.76$), 
$f_\star$ is the stellar mass fraction of the halo,
$f_\rmn{b}$ is the universal baryon fraction ($\Omega_\rmn{b}/\Omega_\rmn{M}$),
$f_\rmn{c}$ is the correction for the baryon depletion in given radial extent,
$m_\rmn{p}$ is the proton mass, and 
$\mu$  is the mean molecular weight for an ionized medium of primordial abundance defined as $\mu = 4/(3X_\rmn{H} + 1 + X_\rmn{H} x_\rmn{e} ) = 0.588$. The quantity $N_{\rmn{e},\Delta}$ is the total number of free electrons in a radius $R_\Delta$, which is analogous to the the total gas mass within $R_\Delta$. We relate $N_{\rmn{e},\Delta}$ to the total mass, $M_\Delta$, with the physical constant above and the variables $f_\star$ and $f_\rmn{c}$. Similarly, Equation \ref{eq:yproj} becomes

\bea
Y &=& \frac{\st}{\me c^2}  \int^{R_{\Delta}}_0 P_\rmn{e} (r) \dd V \nonumber \\ 
&=& \frac{\st}{\me c^2} N_{\rmn{e},\Delta} kT_{\rmn{e},\Delta} \nonumber \\
&=& \frac{\st}{\me c^2} x_\rmn{e} X_\rmn{H} (1 - f_\star)f_\rmn{b}f_\rmn{c} \frac{M_\Delta kT_{\rmn{e},\Delta}}{\mu m_p},
\label{eq:Yobs}
\eea

\noindent here $T_{\rmn{e},\Delta}$ is a characteristic electron temperature of halo, which can be thought of as a virial temperature. Both $\mathcal{T}$ and $Y$ are related to the average optical depth or Compton-y parameter in an aperture through $\langle x \rangle_\Theta = X / (\pi \Theta^2 d^2_A(z))$. Here $x$ is $\te$ or $y$, $X$ is $\mathcal{T}$ or $Y$, and the $\langle x \rangle_\Theta$ refers to the average over an aperture with radial size $\Theta$. The radius $R_\Delta$ is related to $\Theta$ through $R_\Delta = \Theta d_A(z)$ assuming the small angle approximation.

\section{Simulations}
\label{sec:sims}

We compute the observable quantities $\langle \te \rangle_\Theta$ and $\langle y \rangle_\Theta$ from halos generated in simulations of cosmological boxes using a modified version of GADGET-2 \citep{Gadget}, a smoothed particle hydrodynamics (SPH) code. The version of GADGET-2 that we used included sub-grid models for active galactic nuclei (AGN) feedback \citep[for more details see][]{BBPSS}, radiative cooling, star formation, galactic winds, supernova feedback \citep[for more details see][]{SpHr2003}, and cosmic ray physics \citep[for more details see][]{2006MNRAS.367..113P,2007A&A...473...41E,2008A&A...481...33J}. We explored a range of sub-grid models listed in order of increasing complexity:

\begin{itemize}
\item A non-radiative model with only gravitational heating (hereafter {\it non-radiative}) with no star formation.
\item A model with radiative cooling, star formation, galactic winds, supernova feedback, and cosmic ray physics (hereafter {\it radiative cooling}).
\item A model with the addition of AGN feedback to the {\it radiative cooling} model (hereafter {\it AGN feedback}).
\end{itemize}

The {\it non-radiative} model is an extreme ICM model and not presented as a viable alternative to the other models. Furthermore, the {\it non-radiative} model has been shown to be significantly discrepant with cluster observations \citep[e.g.,][]{Puch2008}, but is still shown in this work because it is the simplest sub-grid model and its results are easiest to physically understand.
For each model we ran simulations using 10 different initial conditions.
The physical parameters of these simulations were box sizes of 165 Mpc$/h$, with a resolution of 256$^3$ gas and dark matter (DM) particles. These parameters yield mass resolutions of $M_\rmn{gas} = 3.2\times 10^{9} \rmn{M}_{\odot}/h$ and $M_\rmn{DM} = 1.54\times 10^{10} \rmn{M}_{\odot} /h$. The cosmological parameters used for these simulations were $\Omega_\rmn{M} = \Omega_\rmn{DM} + \Omega_\rmn{b} = 0.25$, $\Omega_\rmn{b} = 0.043$, $\Omega_\Lambda = 0.75$, $H_0=100\,h\,\rmn{km}\,\rmn{s}^{-1}\,\rmn{Mpc}^{-1}$, $h=0.72$, $n_\rmn{s} =0.96$ and $\sigma_8 = 0.8$. These are the cosmological parameters assumed throughout this paper.

We identify halos and calculate their properties in the following way: Halos are initially found using a friends of friends algorithm \citep{Huch1982}; For each identified halo, a center of mass (COM) is computed iteratively followed by the spherical overdensity mass ($M_{\Delta}$) and radius ($R_{\Delta}$). Here $\Delta$ refers to the multiplicative factor applied to the critical density, $\rho_\rmn{cr}(z) \equiv 3 H_0^2 [\Omega_\rmn{M}(1+z)^3 + \Omega_{\Lambda}] / (8\pi\,G)$.
These steps are followed at each redshift slice in the simulations. Then we projected the quantities $\tau$ and $y$ for the 300 most massive clusters from the entire sample of the simulated volumes. Each projection is $8 \times 8$ Mpc$/h$ comoving centered on the cluster's COM and projected down the entire 165 Mpc$/h$ length of the simulation. Finally, we compute $\langle \te \rangle_\Theta$ and $\langle y \rangle_\Theta$ by averaging over the given aperture $\Theta$.

\begin{figure}
\begin{center}
\includegraphics[width=0.99\columnwidth]{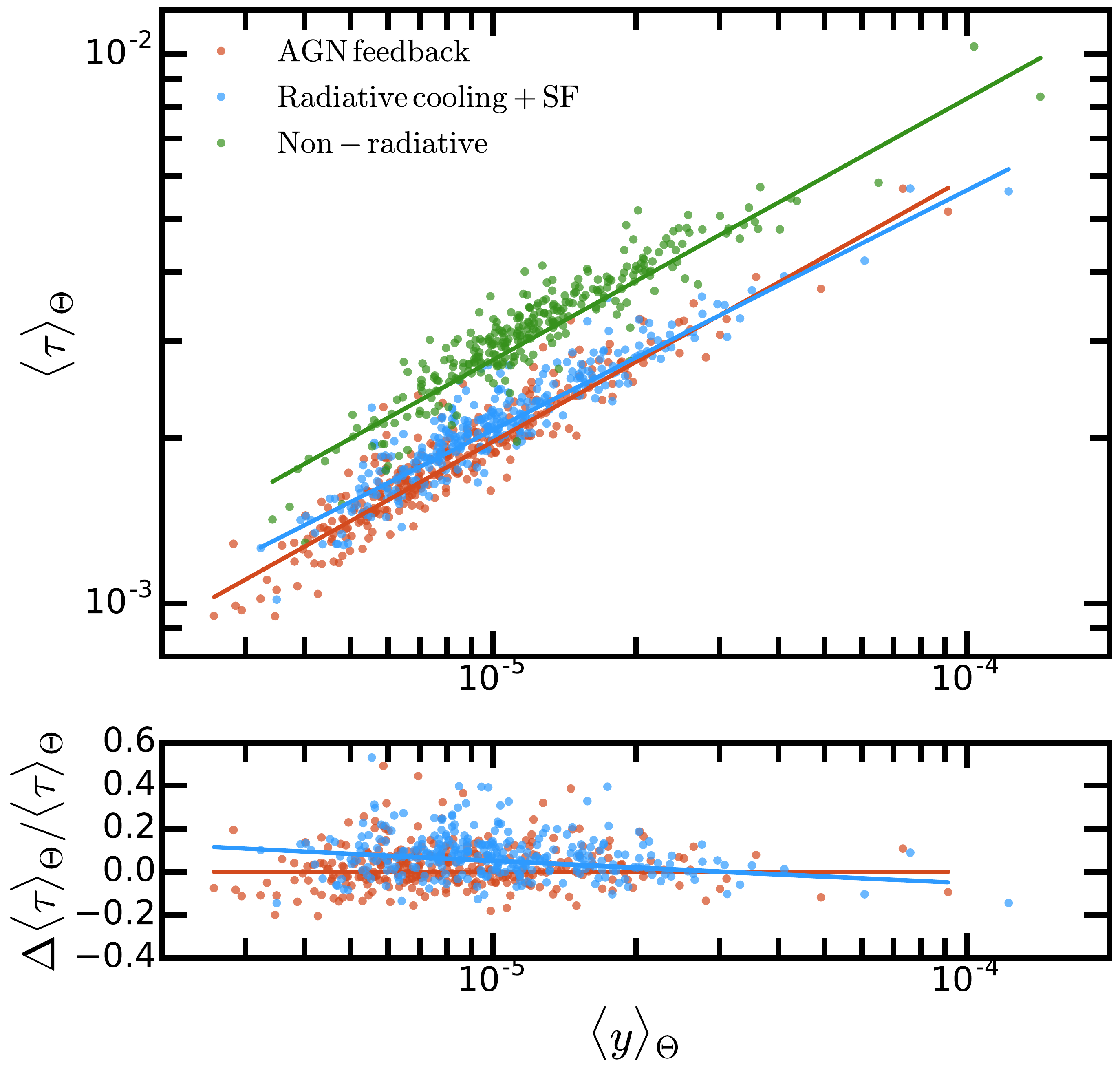}
\caption{$\tauy$ scaling relations for different sub-grid physics models at $z=0.3$ for $\Theta$ =1.3'. Each point represents a simulated halo for a given sub-grid physics model and the corresponding solid lines are the best fit scaling relations. The fractional difference relative to the best fit {\it AGN feedback} parameters are shown in the bottom panel (excluding the {\it Non-radiaive} model). This figure illustrates the tight correlation between $\langle \tau \rangle_\rmn{\Theta}$ and $\langle y \rangle_\rmn{\Theta}$. For the range of masses considered we find on average an 8\% differences between {\it radiative cooling} and {\it AGN feedback} sub-grid physic models (see Section \ref{sec:syst}), which is the largest difference found.
Note that the values on the x and y axis do not include contributions along the line-of-sight beyond the simulated box size in this figure and the subsequent ones.}
\label{fig:tau_y}
\end{center}
\end{figure}

\section{$\te$ relations}
\label{sec:rels}

The optical depth of CMB photons through halos is not directly observable for an individual halo. With tSZ and kSZ observations we measure the pressure ($T_\elct \tau$) and momentum (${\bf \hat{v}_\rmn{p}} \tau$). X-ray observations of the ICM also probe $n_\elct$, since they are proportional to $n_\elct^2 \Lambda(T_\elct)$. However, these X-ray $n_\elct$ profiles are subject to 10\% or greater biases in the de-projection algorithms from either asphericity \citep[e.g.,][]{Vazza2011,Eckert2012}, clumping \citep[e.g.,][]{Simu2011,Morandi2013,Pratt2016}, or metal abundance modeling \citep{Aves2014}. We explore the possibilities of inferring an average optical depth ($\langle \tau \rangle_\rmn{\Theta}$) within an aperture $\Theta$ for a given halo using a power-law scaling relation

\be
\langle \rmn{ln} \tau | A \rangle = \rmn{ln} \tau_0 + m \rmn{ln}(A/A_0)
\label{eq:rel}
\ee

\noindent where $A$ is an observable, $m$ is the power-law slope, $\tau_0$ is the intercept for a given normalization of the observable $A_0$. We choose our fiducial aperture size to be 1.3 arcmins, and explore changing the aperture later in this section. The ideal scaling relation would be independent of sub-grid physics model, have little-to-no scatter, and be related to a direct observable.

\begin{figure}
\begin{center}
\includegraphics[width=0.99\columnwidth]{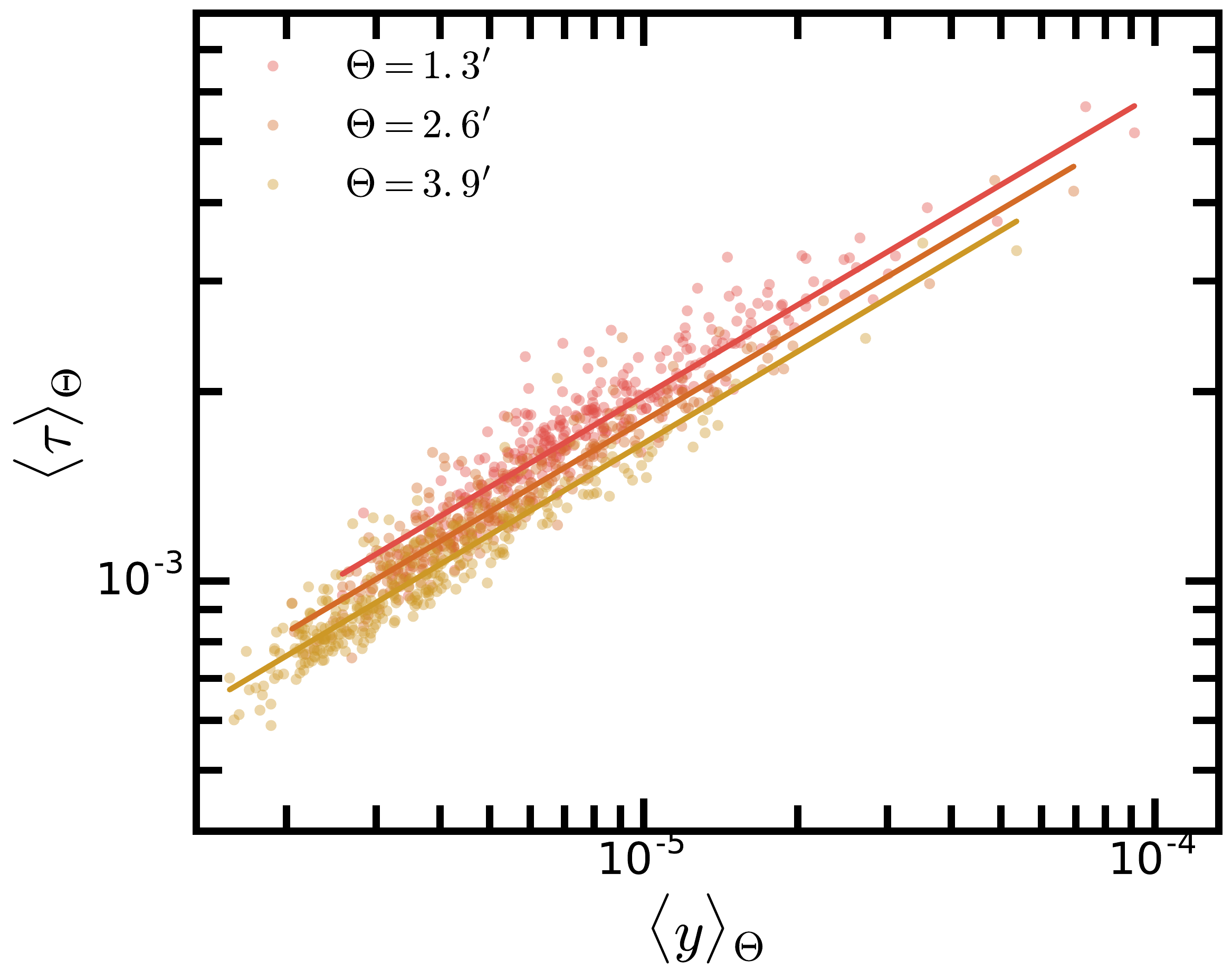}
\caption{$\tauy$ scaling relations for different radial aperture sizes at $z=0.3$ for the {\it AGN feedback} sub-grid model simulations. Each point represents a simulated halo for a given aperture size and the corresponding solid lines are the best fit scaling relations. The tight correlation between $\langle \tau \rangle_\rmn{\Theta}$ and $\langle y \rangle_\rmn{\Theta}$ is preserved when changing the aperture size.}
\label{fig:tau_beam}
\end{center}
\end{figure}

Measurements of $\langle y \rangle_\rmn{\Theta}$ are direct observables for CMB experiments \citep[e.g.,][]{Hand2012,PlankStkLBG,Greco2015,Spacek2016}.
In fact, trivial modifications to certain kSZ estimators are needed to measure $\langle y \rangle_\rmn{\Theta}$ \citep[e.g.,][]{Schaan2016}.
In Figure \ref{fig:tau_y} we show the relation between $\langle \tau \rangle_\rmn{\Theta}$ and $\langle y \rangle_\rmn{\Theta}$ for the three simulated sub-grid models (hereafter, we refer to this relation as $\tauy$) in a $\Theta=1.3'$ aperture. The {\it non-radiative} model shows a significant difference from the other two models, which results from this model not forming stars. The bottom panel of Figure \ref{fig:tau_y} shows the relative differences between the best fit {\it AGN feedback} relation and the {\it AGN feedback} and {\it radiative cooling} simulations. These are small, which is encouraging for the viability of the $\langle y \rangle_\rmn{\Theta}$ relation. The maximum differences between the mean $\tauy$ relations of the {\it radiative cooling} and {\it AGN feedback} models have $\Delta \ln \te_0 \leqslant 0.03$ and  $\Delta m \leqslant 0.05$ (see Table \ref{tab:rels} and caption for percent differences). The small differences between the $\tauy$ relations for {\it radiative cooling} and {\it AGN feedback} models are non-trivial since they both have different gas fraction properties \citep{BBPS3} and $Y-M$ scaling relations \citep{BBPS1}. The scatter about the $\tauy$ relations ranges from $\sigma_{\rmn{ln}\tau}$ = 0.079 to 0.096 for $\Theta = 1.3'$ and the scatter gets larger as we increase the aperture (see Figure \ref{fig:tau_beam}) for the redshifts $z=0.5$ and $z=0.7$. At $z=0.3$ the $\Theta=1.3'$ aperture samples the interior of the halos and is subject to larger variance between halos. The scatter found in these relations are always smaller than the overall $\Delta \tau / \tau$ over the same sample of clusters with a mass range of $M_{500} = 10^{14}-10^{15} M_\odot$.

Using Equations \ref{eq:tauobs} and \ref{eq:Yobs} we have a theoretical understanding for this $\tauy$ scaling relation. We find that
\be
\mathcal{T} = \frac{\me c^2}{kT_{\rmn{e},\Delta}} Y,
\ee

\noindent and $\langle \tau | y \rangle \propto 1 /(T_{\rmn{e},\Delta})$. Thus, the $\tauy$ relation and its scatter reflect the difference in $1/(T_{\rmn{e},\Delta})$ at fixed total gas pressure. This low-scatter correlation in the $\tauy$ relation is not surprising given that the Planck satellite \citep{PlankSZcalib2011} and the Weighing the Giants collaboration \citep{Mantz2016} found tight correlations between $Y$ and $M_\rmn{gas}$. Previous X-ray measurements of clusters also showed a strong, low-scatter correlation between $T_\rmn{X}$ and $M_\rmn{gas}$  \citep[e.g.,][]{Mohr1999,VVP2002,Kravt2006}. While the $\tauy$ and $T_\rmn{X} - M_\rmn{gas}$ relations are not exactly the same, all these observations further support that at fixed gas pressure the scatter in the gas-mass weighed temperature should be small. 

\begin{figure}
\begin{center}
\includegraphics[width=0.99\columnwidth]{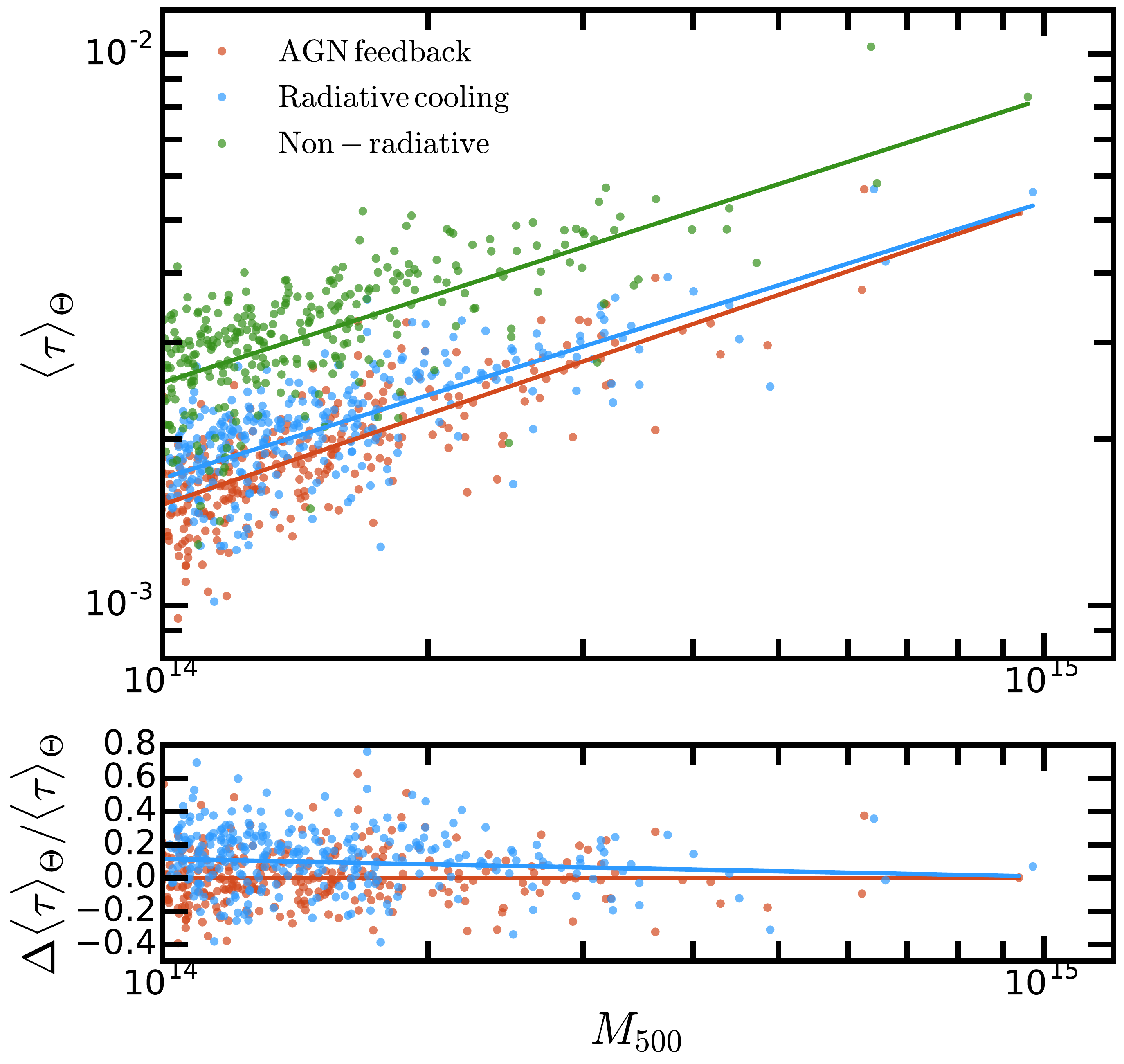}
\caption{$\tauM$ scaling relations for different sub-grid physics models at $z=0.3$ for $\Theta$ =1.3'.
 Each point represents a simulated halo for a given sub-grid physics model and the corresponding solid lines are the best fit scaling relations.
 The fractional difference relative to the best fit {\it AGN feedback} parameters are shown in the bottom panel (excluding the {\it Non-radiaive} model).
 In comparison to the $\tauy$ relation (see Figure \ref{fig:tau_y}) the $\tauM$ relation has larger differences between {\it radiative cooling} and {\it AGN feedback} sub-grid physics models and larger scatter.}
\label{fig:tau_m}
\end{center}
\end{figure}

An obvious scaling relation to investigate as well is the $\tauM$ relation (here $M$ is the $M_{500}$ of the halo). This relation directly follows from the self-similar model \citep{Kaiser1986} where $\langle \tau \rangle$ should be a function of halo mass that scales like $f_\rmn{gas} = (\Omega_\rmn{b} - \Omega_\star)/ \Omega_\rmn{m}$. In the top panel of Figure \ref{fig:tau_m} we show the relation for various sub-grid physics models. The mean $\tauM$ relations are shown with the solid lines. The bottom panel of Figure \ref{fig:tau_m} shows the relative differences between the best fit {\it AGN feedback} relation and the {\it AGN feedback} and {\it radiative cooling} simulations. Ignoring the {\it non-radiative} model, we find that the $\tauM$ relation parameters are more dependent on the sub-grid physics model than the $\tauy$ relation. The fact that the $\tauM$ relation parameters are more sensitive to sub-grid physics models is analogous to previous studies on $f_\rmn{gas} (M)$ \citep[e.g.,][]{2005ApJ...625..588K,2006MNRAS.365.1021E,2008ApJ...687L..53P,Fabjan2010,McCarthy2011,2011MNRAS.413..691Y,BBPS3,Plan2013,Semb2013,Pike2014,LeBM2014,McCarthy2016,Troung2016} and is not surprising.

\begin{figure}
\begin{center}
\includegraphics[width=0.99\columnwidth]{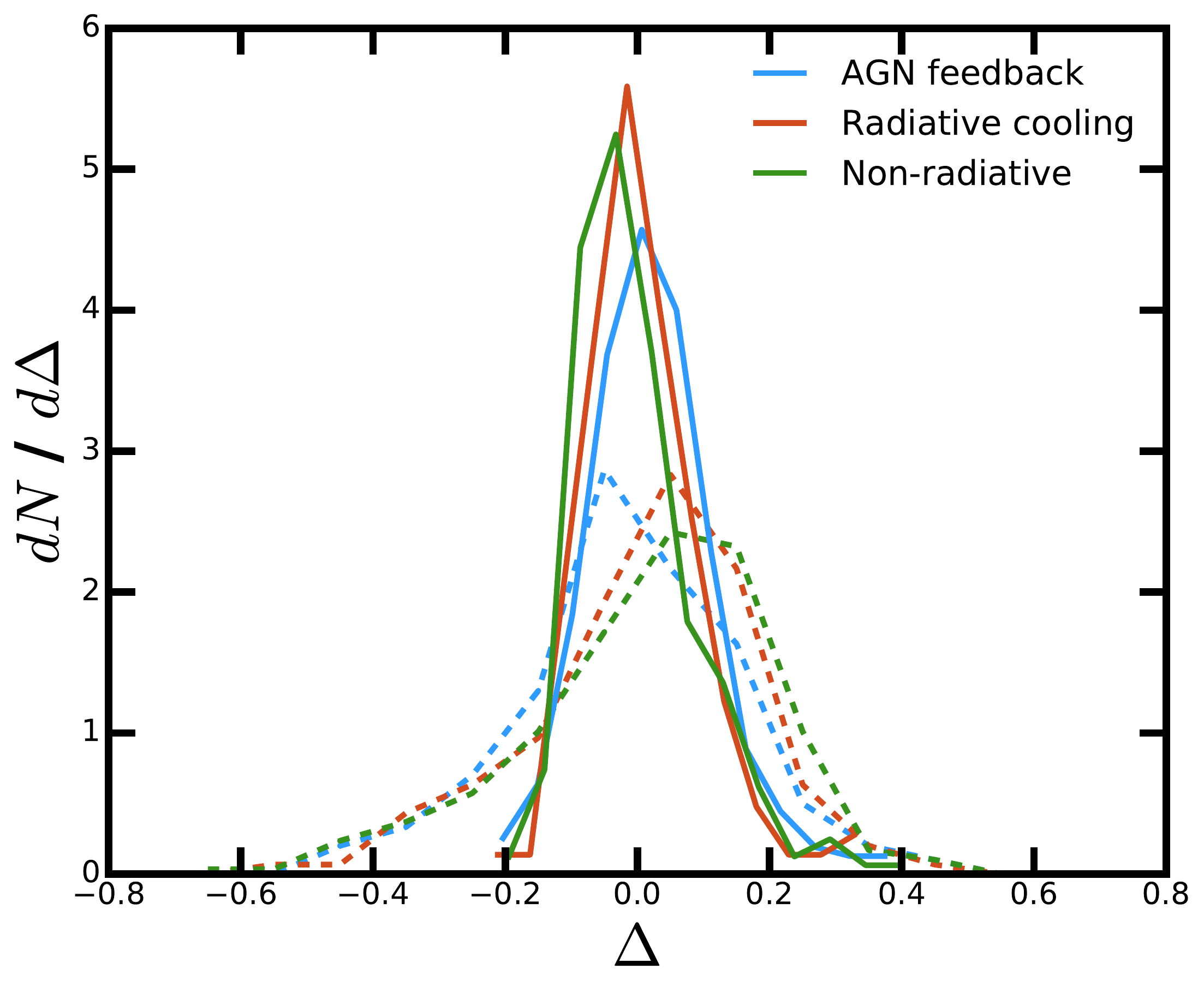}
\caption{Probability distribution functions of the residuals of the best fit scaling relations for $\tauy$ (solid lines) and $\tauM$ (dashed lines) at $z=0.3$ for $\Theta$ = 1.3'. 
Here $\Delta$ is defined as difference between the natural logarithms for individual halo tau values and the best relations.
The residuals from the $\tauy$ relation have a narrow distribution than the $\tauM$ relation independent of the sub-grid physics model.}
\label{fig:pdf_comp}
\end{center}
\end{figure}

Equation \ref{eq:tauobs} provides the theoretical basis for understanding the $\tauM$ scaling relation. This equation shows that $\langle \tau | M \rangle  \propto (1- f_\star)f_\rmn{c}$.
Thus, the $\tauM$ relation and its scatter reflect the difference in $f_\star$ and $f_\rmn{c}$ at fixed total mass. We directly compare the probability distribution function (PDF) of the residuals about the best fit $\tauy$ and $\tauM$ relations in Figure \ref{fig:pdf_comp} at $z=0.3$ and $\Theta = 1.3'$. The scatter about these relations is larger for $\tauM$ than for $\tauy$ (see Figure \ref{fig:pdf_comp} and Table \ref{tab:rels}). Two clear advantages to using the $\tauy$ relation over the $\tauM$ relation are the quantity $\langle y \rangle_\rmn{\Theta}$ is a direct observable whereas halo mass is not, and the scatter and dependence on the sub-grid physics model are much smaller for $\tauy$ than for $\tauM$.

\subsection{Systematic Uncertainties in the $\te$ relations}
\label{sec:syst}

The best fit $\tauy$ and $\tauM$ relation parameters in Table \ref{tab:rels} do not have uncertainties quoted with them.
We found errors from the fitting process are insignificant compared to the differences between the parameters of the {\it radiative cooling} and {\it AGN feedback} sub-grid physics models.
These difference are what we choose to quote as systematic uncertainties to the $\tauy$ and $\tauM$ relation and its parameters.
The two sub-grid models used here do not begin to explore the vast parameter space of sub-grid models for the ICM. However, they reasonably bracket two vastly different sub-grid models, with and without AGN feedback.
For this reason, we conservatively chose the largest differences between these models as the uncertainty.
We quote the systematic uncertainties on the parameters at each aperture in Table \ref{tab:rels} and it is these uncertainties that should be used when extrapolating these relations.
We find that the systematic uncertainties on the parameters get smaller as the aperture increases, which illustrates that the average gas-mass weighted temperature of halos at a fixed gas mass becomes less sensitive to the details of sub-grid modeling as one probes further outside the halo.
The other way that we estimate the systematic errors on the $\tauy$ and $\tauM$ relations is to calculate the percent difference on $\langle \tau \rangle_\rmn{\Theta}$ assuming the {\it AGN feedback} relations for the halos from the {\it radiative cooling} simulations. We refer to these systematic errors as $\systa$ in Table \ref{tab:rels}.
A feature of this approach is that it is independent of the the choice for $A_0$ from Equation \ref{eq:rel}, but it is only valid over the mass ranges considered in the work.
Beyond the masses considered here one should propagate the systematic errors on the individual parameters.
The trends found for the individual parameters are the same for $\systa$.

\begin{table*}
  \caption[Mass and redshift dependencies of the average thermal
  pressure profile]{Best fit scaling relation parameters for the $\tauy$ and $\tauM$ relations.}
  \label{tab:rels}
  \begin{center}
   \leavevmode
   \begin{tabular}{l|lccc|ccc|ccc} 
     \hline \hline              
     \multicolumn{2}{c}{ } & \multicolumn{3}{c|}{AGN feedback} &  \multicolumn{3}{c|}{Radiative Cooling} & \multicolumn{3}{c}{Shock heating}\\
     \hline 
    &  & $z=0.3$ & $z=0.5$ & $z=0.7$ &  $z=0.3$ & $z=0.5$ & $z=0.7$ &  $z=0.3$ & $z=0.5$ & $z=0.7$\\
     \hline
     $\tauy$ &ln$\tau_0$   &-6.23 & -6.34 & -6.45  &   -6.18 & -6.29 & -6.45 &  -5.88 & -6.06 & -6.20\\ 
     $\Theta$ = 1.3' &$m$   & 0.48 & 0.49 & 0.49 & 0.44 & 0.47 & 0.44 & 0.48 & 0.46 & 0.48\\        
     &$\sigma_{\rmn{ln}\tau}$ & 0.096 & 0.087 & 0.096 & 0.089 & 0.079 & 0.087 & 0.096 & 0.073 & 0.082\\ 
     &$\systa$ & 8\% & 6\% & 7\% & - & - & - & - & - & -\\ 
     \hline
     &ln$\tau_0$   &-6.27 & -6.40 & -6.53  &   -6.23 & -6.37 & -6.55 &  -5.97 & -6.16 & -6.33\\ 
     $\Theta$ = 1.8' &$m$   & 0.48 & 0.49 & 0.48 & 0.44 & 0.47 & 0.44 & 0.48 & 0.47 & 0.47\\        
     &$\sigma_{\rmn{ln}\tau}$ & 0.090 & 0.086 & 0.10  & 0.083 & 0.080 &  0.095 & 0.080 & 0.074 & 0.093\\ 
     &$\systa$ & 6\% & 5\% & 6\% & - & - & - & - & - & -\\ 
     \hline
      &ln$\tau_0$   &-6.32 & -6.47 & -6.64 &   -6.30 & -6.46 & -6.65 &  -6.07 & -6.28 & -6.47\\ 
     $\Theta$ = 2.6' &$m$   & 0.48 & 0.49 & 0.46 & 0.45 & 0.48 & 0.44 & 0.47 & 0.49 & 0.47\\        
     &$\sigma_{\rmn{ln}\tau}$ & 0.087 & 0.093& 0.11 & 0.080 & 0.089 & 0.11 & 0.072 & 0.087 & 0.11\\ 
     &$\systa$ & 4\% & 3\% & 4\% & - & - & - & - & - & -\\ 
     \hline
      &ln$\tau_0$   &-6.40 & -6.57 & -6.79  &   -6.41 & -6.55 & -6.78 &  -6.22 & -6.41 & -6.62\\ 
     $\Theta$ = 3.9' &$m$   & 0.48 & 0.49 & 0.44 & 0.45 & 0.49 & 0.43 & 0.47 & 0.50 & 0.45\\        
     &$\sigma_{\rmn{ln}\tau}$ & 0.091 & 0.11 & 0.13 & 0.087 & 0.11 & 0.13 & 0.081 & 0.11 & 0.13\\
     &$\systa$ & 2\% & 2\% & 2\% & - & - & - & - & - & -\\  
     \hline
     \hline
     $\tauM$ &ln$\tau_0$   &-6.49 & -6.83 & -7.08  &   -6.38 & -6.76 & -6.98 &  -5.98 & -6.40 & -6.66\\ 
     $\Theta$ = 1.3' &$m$   & 0.54 & 0.60 & 0.56 & 0.50 & 0.60 & 0.48 & 0.51 & 0.56 & 0.53\\        
     &$\sigma_{\rmn{ln}\tau}$ & 0.17 & 0.14 & 0.15 & 0.17 & 0.13 & 0.12 & 0.20 & 0.14 & 0.12\\ 
     &$\systa$ & 12\% & 12\% & 11\% & - & - & - & - & - & -\\ 
     \hline
   \end{tabular}
  \end{center}
  \begin{quote}
    \noindent 
    The systematic uncertainties for each parameter are estimated by the largest relative differences between the {\it radiative cooling} and {\it AGN feedback} sub-grid physics models for each aperture across all redshifts. For $\Theta$ = 1.3' the systematic uncertainty on ln$\tau_0$ is 5\% and on $m$ is 8\% in the $\tauy$ relation. For $\Theta$ = 1.8' the systematic uncertainty on ln$\tau_0$ is 4\% and on $m$ is 8\% in the $\tauy$ relation. For $\Theta$ = 2.6' and $\Theta$ = 3.9' $\tauy$ relations the systematic uncertainties on ln$\tau_0$ and $m$ are 2\% and 6\%, respectively. For $\Theta$ = 1.3' $\tauM$ relation the systematic uncertainty on ln$\tau_0$ is 12\% and on $m$ is 14\%.
  \end{quote}
\end{table*}

\section{Discussion}
\label{sec:disc}

There is a systematic error floor in the approach we present here that results from the uncertainties associated with the sub-grid models in simulations.
Observational calibrations of the $\tauy$ and $\tauM$ relations have the potential to mitigate this systematic floor. An initial demonstration of such a calibration is shown in \citet{FDB2016}. One possible future observational method would be to use the statistical estimator from \citet{DS2009} to calibrate $\tau$ from galaxy or clusters through cross correlations. This follows a similar proposal to measure the fluctuations in $\tau$ from reionzation by cross-correlating with measurements from 21cm experiments \citep{Meerburg2013}. Here the amplitude of cross spectra proposed are directly proportional to the amplitude of the $\tauy$ and $\tauM$ relations. Thus, a highly significant measurement of these cross spectra would empirically calibrate the $\tauy$ and $\tauM$ relations and eliminate the systematic uncertainties from astrophysics that are not captured by current simulations.

Proposed CMB spectral distortion satellites, like PIXIE \citep{PIXIE2011}, will constraint the total thermal energy of electrons in the observable Universe at the sub-percent level and measure the sky-averaged relativistic tSZ signal \citep{Hill2015}. Thus, it will allow us to determine global properties of ionized gas to high precision and impose an {\it integral constraint} on models for energetic feedback processes and galaxy formation. Future X-ray satellites, like eROSITA \citep{Erosita2012}, have the potential to measure the ionized gas masses for all the halos above a few $\times 10^{13} M_\odot$ across the entire sky. Given such a large sample size, stacking techniques will alleviate some of the systematic errors associated with asphericity for future eROSITA measurements, but not the systematics associated with clumping.

Recently, \citet{Sugiyama2016} proposed to marginalize over the amplitude information in the pairwise momentum estimator, i.e. over the halo optical depth, and constrain cosmology from the scale-dependence of the Legendre expansion of the pairwise momentum. In combination with galaxy clustering, this kSZ {\it shape} information helps constrain the expansion and growth rate of structure in the Universe. Our approach is complementary to theirs, as it instead focuses on the amplitude information in the signal, through the $\tauy$ and $\tauM$ relations. This should allow to recover the information lost when marginalizing over the amplitude of the kSZ signal.

\section{Conclusion}
\label{sec:con}

The recent kSZ detections via cross-correlations have renewed the interest in using these measurements as cosmological probes.
Breaking the degeneracy between the optical depth and peculiar velocity is paramount as we approach high-precision kSZ measurements and attempt to constrain cosmological parameters with such measurements.
In this work we proposed to use scaling relations to infer the optical depth of halos, thus breaking the degeneracy.

We used cosmological hydrodynamic simulations to explore two possible scaling relations, $\tauy$ and $\tauM$.
We found that the $\tauy$ relation has many advantages over the $\tauM$ relation:
\begin{enumerate}
\item The observable quantity in the $\tauy$ relation is the tSZ signal within a physical aperture where as the $\tauM$ relation requires halo masses which are not direct observables and have additional uncertainties associated with their inference.
\item The $\tauy$ relation depends less on the sub-grid physics model compared to the  $\tauM$ relation (See Figures \ref{fig:tau_y} and \ref{fig:tau_m} and Table \ref{tab:rels}). For example, the percent differences on the inferred $\langle \tau \rangle_\rmn{\Theta}$ values when using the incorrect scaling relation for a different sub-grid model, $\systa$, are 50\% larger for $\tauM$ compared to $\tauy$. The increase in systematic uncertainties on scaling relation parameters $\tau_0$ and $m$ are roughly 2$\times$ larger (140\% and 75\% to be exact, respectively) for $\tauM$ compared to $\tauy$.
\item The intrinsic scatter in the $\tauy$ relation is less than the $\tauM$ relation (See Figure \ref{fig:pdf_comp} and Table \ref{tab:rels}). This result is in agreement with X-ray observations that find a low-scatter correlation between $T_\rmn{X}$ and $M_\rmn{gas}$.
\end{enumerate}

The scaling relations we present in this work are based on simulations and they are subject to the systematic uncertainties associated the sub-grid physics models in these simulations. We calculated these systematic uncertainties to be less than 10 percent (See Table \ref{tab:rels} for the exact numbers) and they are always smaller than the intrinsic scatter. We use the largest differences between two vastly different sub-grid models, with and without AGN feedback, to estimates these systematic uncertainties. Note that the sub-grid {\it AGN feedback} model we used has subsequently been found to agree with low and intermediate redshift measurements of high-mass cluster pressure profiles \citep{PlnkP2013,Mcdn2014} and is consistent with measurements of the stellar and gas content in low-redshift clusters \citep{BBPS3}. Therefore, the difference between our sub-grid models should yield conservative estimates despite the fact that they do not fully explore the parameter space of sub-grid models for the ICM.

Future observations could potentially empirically constrain the $\tauy$ relation and eliminate systematic uncertainties. The observations include cross-correlations between halos and reconstructed optical depth using the statistical estimator from \citet{DS2009}, CMB spectral distortion measurements, and observations from future X-ray satellites, like eROSITA. We leave the detailed forecasts of these prospective observations for future work.

\acknowledgments

We thank Emmanuel Schaan, Simone Ferraro, Michael Niemack, David Spergel, Edward Wollack, and Bruce Partridge for their productive discussions and comments on this work.
NB acknowledges the support from the Lyman Spitzer Fellowship.

\bibliography{nab}
\bibliographystyle{apj}

\begin{appendix}

\section{Density Profile}
\label{sec:fit} 

Theoretical predictions for the optical depth of clusters and groups can be directly calculated from the gas density profile (see Section \ref{sec:thry}).
For massive cluster there exist observational gas density profiles measured by X-ray satellites out to $\sim R_{500}$ \citep[e.g.,][]{Vik2006,2008MNRAS.383..879A,Croston2008,McDn2013,Pratt2016} with the exception of a couple nearby X-ray bright clusters that have observations beyond $R_{500}$ \citep[e.g.,][]{Simu2011}. In cross-correlation the combinations X-ray and tSZ observations have yielded stacked gas density profiles that go beyond $R_{500}$ \citep[e.g.,][]{PlnkP2013,Eckert2012}, for a sub-set of massive clusters. Beyond $R_{500}$ there are theoretical gas density profiles from analytic calculations \citep[e.g.,][]{KS2001,OBB2005,Patej2015} and predictions from numerical simulations \citep[e.g.,][]{Ronc2006,Nagai2007,Vazza2010,Pike2014}, with some of simulations provided a fitting function to their results. Like most halo profiles simulations show that the gas density profile is close to self-similar, but recently it was shown that at large radii self-similarity is broken by the mass accretion history of the halo \citep{Lau2015}.

Here we provide a parametric fitting function for the density profile as a function of halo mass and redshift that is calibrated from simulations. The stacked average profiles $\bar{\rho}_{\rmn{gas}} = \rho_\rmn{crit}(z) \bar{\rho}_{\rmn{fit}}$ are fit to a restricted version of the generalized NFW profile \citep{Zhao1996}.

\begin{equation}\label{eq:rhofit} 
\bar{\rho}_{\rmn{fit}} = \rho_0
   \left(x/x_{\rmn{c}}\right)^{\gamma}\left[1 +
   \left(x/x_{\rmn{c}}\right)^{\alpha} \right]^{- \left(\frac{\beta - \gamma}{\alpha}\right)}, \ x \equiv r/R_{\Delta},
\end{equation}

\noindent where the fit parameters are an intermediate slope $\alpha$, an amplitude
$\rho_0$ and a power law index $\beta$ for the asymptotic fall off of the profile.
There is substantial degeneracy between fit parameters, so we fix the core-scale $x_{\rmn{c}} = 0.5$
and $\gamma = -0.2$. We find the best-fit parameters using a Markov Chain Monte Carlo (MCMC) algorithm from \citet{Emcee2013} that finds the minimum $\chi^2$. 

We fit for deviations from the self-similar mass and redshift scaling and treat each parameter as a
separable function of mass and redshift. The fit parameters are constrained to be of the following form:  For generic
parameter $A$, we have
\begin{equation}\label{eq:rho_mfit}
  A = A_0\left(\frac{M_{\Delta}}{10^{14}
    \,\mathrm{M}_{\sun}}\right)^{\alpha_{\mathrm{m}}} \left(1 + z\right)^{\alpha_{\mathrm{z}}}.
\end{equation}

\begin{figure*}
 \begin{minipage}[t]{0.5\hsize}
    \centering{\small{\em Mass}}
  \end{minipage}
  \begin{minipage}[t]{0.5\hsize}
    \centering{\small{\em Redshift}}
  \end{minipage}
\begin{center}
  \hfill
  \resizebox{0.50\hsize}{!}{\includegraphics[bb=7 7 560 537]{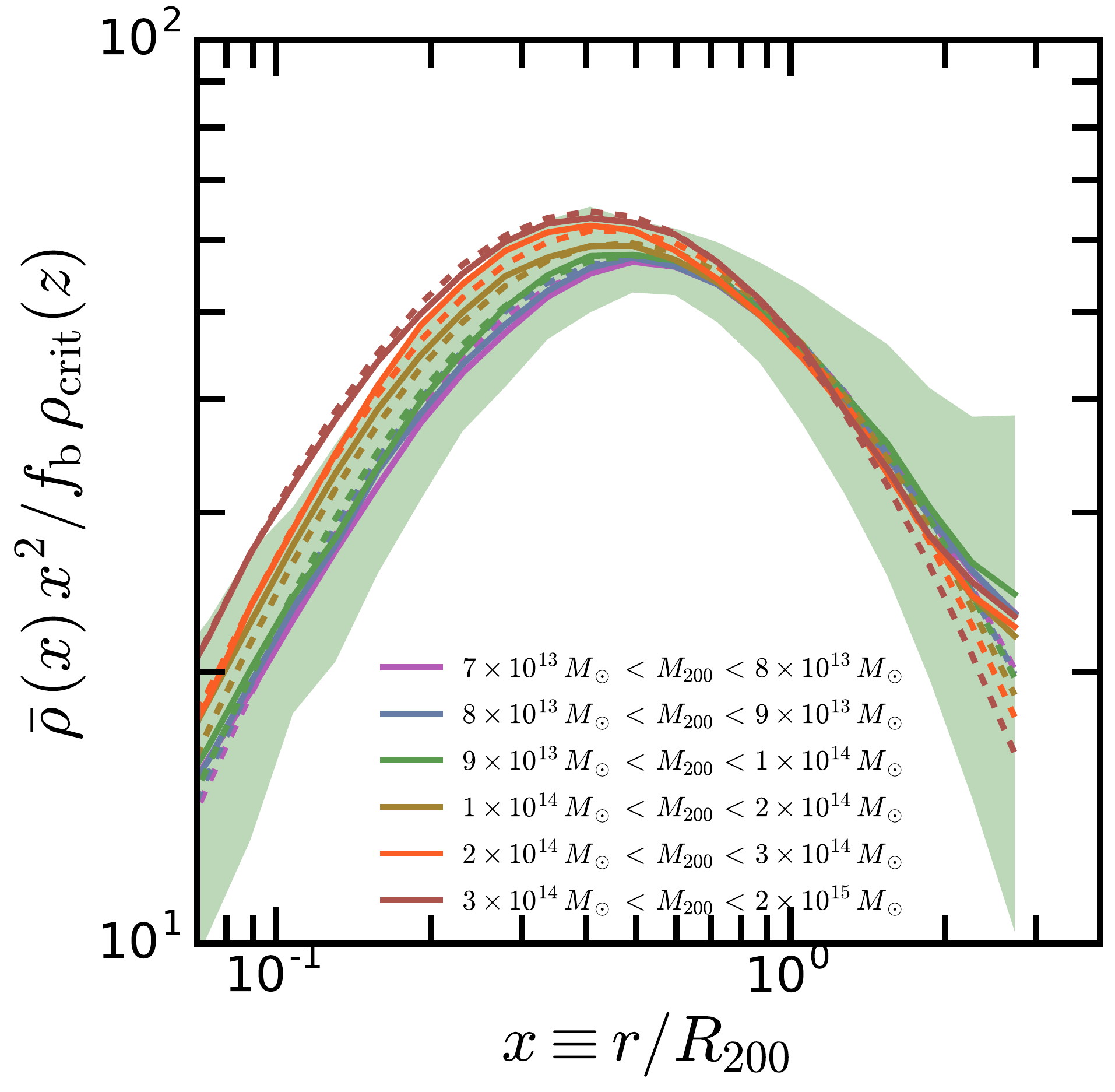}}
  \resizebox{0.50\hsize}{!}{\includegraphics[bb=7 7 560 537]{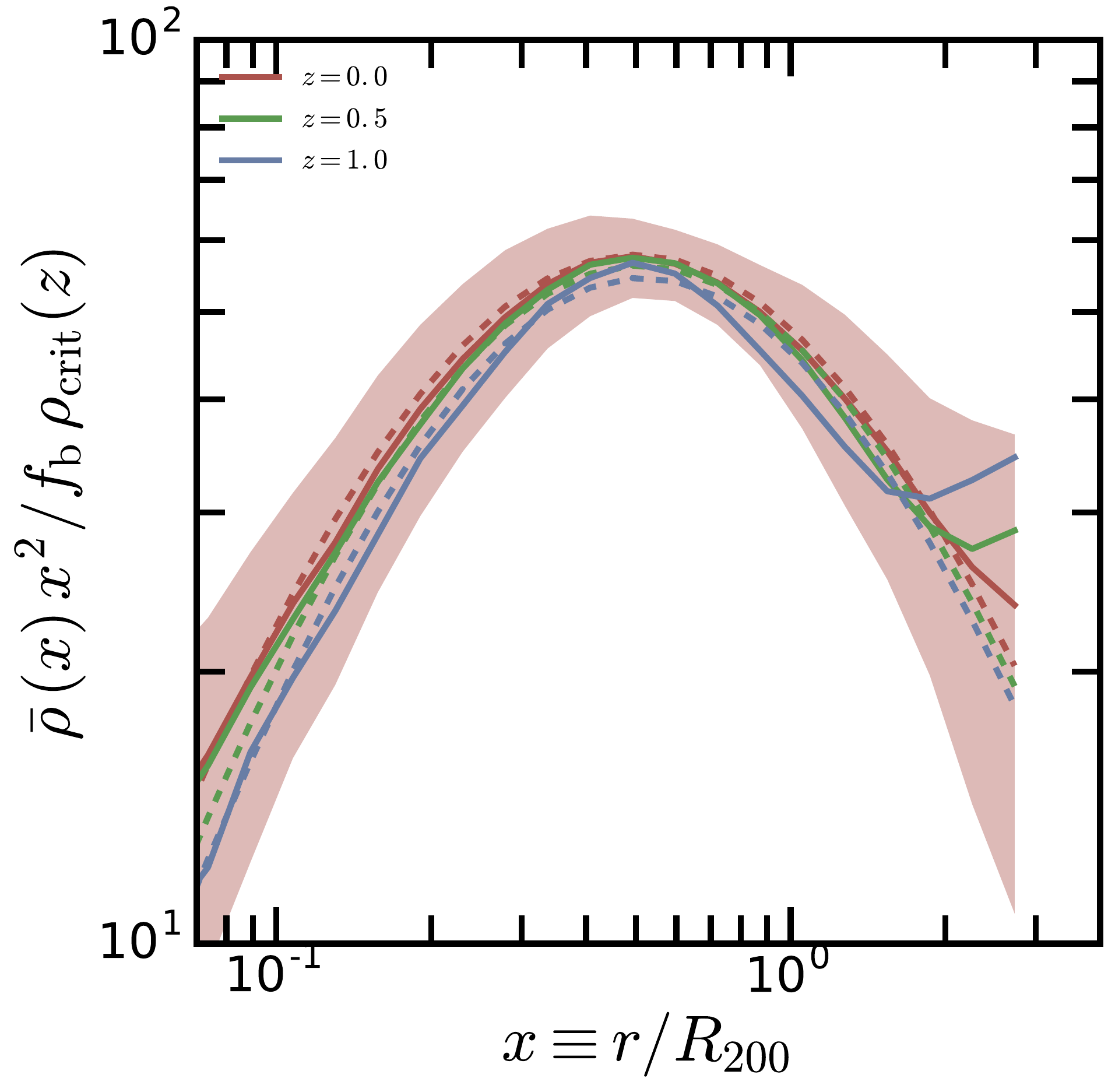}}\\
\end{center}
\caption{Normalized average gas density profiles and parameterized fits to these profiles from simulations with AGN feedback scaled by $(r/R_{200})^2$, in mass bins (left panel) and redshift bins (right panel). Here, we have independently fit each mass and redshift bin. The bands show the standard deviation of the average cluster in the $9\times10^{13} M_\odot < M_{200} < 1\times10^{14} M_\odot$ bin (left) and lowest redshift bin (right). The generalized NFW profile with fixed $x_{\rmn{c}}$ and $\gamma$ fits the average profiles well in the majority of the mass and redshift bins. The upturns at large radii are due to contributions from nearby clusters and substructure.}
\label{fig:rhoprof}
\end{figure*}

We show the average halo gas density profile (scaled by $x^2$) in Figure \ref{fig:rhoprof} with the corresponding parameterized fits to these profiles and a function of mass (left panel) and redshift (right panel).
The fitting function from Equations \ref{eq:rhofit} and \ref{eq:rho_mfit} and parameters in Table \ref{tab:mzfit} provide an accurate fit over all mass and redshift ranges, which are well within the variance about these relations. 
At large radii, especially at higher redshift, is where we find the largest deviation from the fitting function. This deviation happens at location where the contributions of other nearby halos and sub-halos, the 2-halo term, becomes important and its exact location for each halo will depend on the mass accretion history of the halo \citep{Lau2015}.

\begin{table*}[t]
  \caption[Mass and redshift dependencies of the average thermal
  pressure profile]{Mass and Redshift Fit Parameters for
  Eqns.~(\ref{eq:rhofit})  and (\ref{eq:rho_mfit}).}
  \label{tab:mzfit}
  \begin{center}
   \leavevmode
   \begin{tabular}{lccc|ccc} 
     \hline \hline              
     & \multicolumn{3}{c}{AGN feedback $\Delta = 200$} & \multicolumn{3}{c}{Shock heating $\Delta = 200$}\\
     \hline 
     Parameter & $A_{\mathrm{m}} = A_{\mathrm{z}}$ & $\alpha_{\mathrm{m}}$ & $\alpha_{\mathrm{z}}$ & $A_{\mathrm{m}} = A_{\mathrm{z}}$ & $\alpha_{\mathrm{m}}$ & $\alpha_{\mathrm{z}}$\\
     \hline
     $\rho_0$   & 4.0$\times10^{3}$ & 0.29 &-0.66 &   1.9$\times10^{4}$ & 0.09 & -0.95\\ 
     $\alpha$   & 0.88 & -0.03 & 0.19& 0.70 & -0.017 & 0.27 \\        
     $\beta$ & 3.83 & 0.04 & -0.025 & 4.43 & 0.005 & 0.037 \\ 
     \hline
   \end{tabular}
  \end{center}
  \begin{quote}
    \noindent 
    The input weights are chosen to be the inverse variances of fit
    parameter values from the individual gas density fits for each cluster
    within the bin.
  \end{quote}
\end{table*}

\end{appendix}

\end{document}